\documentclass[aps,twocolumn,showpacs]{revtex4}
\usepackage{epsfig}
\usepackage{amsbsy}
\usepackage{amsmath}
\usepackage{amsfonts}
\usepackage{graphicx}
\usepackage{latexsym}

\newcommand{\rf}[1] {(\ref{#1})}

\begin{document}

\title{Spin squeezing and pairwise entanglement for symmetric multiqubit states}
\author{Xiaoguang Wang and Barry C Sanders}
\affiliation{Department of Physics and Australian Centre of Excellence for Quantum Computer Technology, \\
Macquarie University, Sydney, New South Wales 2109, Australia.}

\date{\today}

\begin{abstract}
We show that spin squeezing implies pairwise entanglement for arbitrary symmetric multiqubit states. If the squeezing parameter is less than or equal  to 1, we demonstrate a quantitative relation between the squeezing parameter and the concurrence for the even and odd states.  We prove that the even states generated from the initial state with all qubits being spin down, via the one-axis twisting Hamiltonian, are spin squeezed if and only if they are pairwise entangled. For the states generated via the one-axis twisting Hamiltonian with an external transverse field for any number of qubits greater than 1 or via the two-axis counter-twisting Hamiltonian for any even number of qubits, the numerical results suggest that such states are spin squeezed if and only if they are pairwise entangled.
\end{abstract}
\maketitle

\section{Introduction}
Spin squeezed states~\cite
{Kitagawa,Wineland,Agarwal,Lukin,Vernac,Kuzmichqnd,Sorensen1,Youli,hald,kasevich,uffe,Dominic,Usha,Usha1,Gasenzer,Stockton,Add1,Add2,WangSpin,WangSuper} are quantum correlated states with reduced
fluctuations in one of the collective spin components, with possible applications in atomic interferometers and high precision atomic
clocks.
It is found that spin squeezing
is closely related to and implies quantum entanglement~\cite{Sorensen01,Dominic2,Sanmore}. 
As there are various kinds of entanglement, a
question naturally arises: what kind of entanglement does spin squeezing imply?
Recently, it has been found that, for a two-qubit symmetric state, spin
squeezing is equivalent to its bipartite entanglement~\cite{Kitagawa2}; i.e., spin squeezing implies bipartite entanglement and vice versa. Here, we generalize the above result to the multiqubit case, and study relationships between spin squeezing and quantum entanglement.

Specifically, we first show that spin squeezing implies pairwise entanglement for arbitrary symmetric multiqubit states. If the squeezing parameter $\xi^2\le 1$ (defined below), we give a quantitative relation between the squeezing parameter and the concurrence~\cite{Conc} for even and odd states, where the concurrence is a measure of the degree of two-qubit entanglement, and even (odd) states refer to those where only even (odd) excitations contribute.  
We further consider the multiqubit states dynamically 
generated from the initial state with all qubits being spin down via (i) the one-axis twisting Hamiltonian~\cite{Barry89,Kitagawa}, (ii) the one-axis twisting Hamiltonian with an external transverse field~\cite{Law}, and (iii) the two-axis counter-twisting Hamiltonian~\cite{Kitagawa}. We prove that the states generated via the first Hamiltonian are spin squeezed if and only they are pairwise entangled. For the states generated via the second Hamiltonian and third Hamiltonian with even number of qubits, numerical results for the squeezing parameter and concurrence suggest that the spin squeezing implies pairwise entanglement and vice versa.

\section{Spin squeezing and pairwise entanglement}
A collection of $N$ qubits is represented by the collective operators 
\begin{equation}
S_\alpha =\sum_{i=1}^N\frac{\sigma _{i\alpha }}{2},\quad \alpha \in \{x,y,z\},
\end{equation}
where $\sigma _{i\alpha}$ are the Pauli operators for the $i^{\text{th}}$
qubit. The collective operators satisfy the usual angular momentum 
commutation relations.  
Following Kitagawa and Ueda's criterion of spin squeezing, we
introduce the spin squeezing parameter~\cite{Kitagawa}
\begin{equation}
\xi^2=\frac{2\left(\Delta S_{{\vec n}_\perp}\right)^2}{J}=\frac{4\left( \Delta S_{{\vec n}_\perp}\right)^2}{N},  \label{eq:squeeze}
\end{equation}
where the subscript ${\vec n}_\perp $ refers to an axis perpendicular to the mean spin $\langle{\vec S}\rangle$, where the minimal value of the variance $(\Delta S)^2$ is obtained, $J=N/2$, and $S_{{\vec n}_\perp}={\vec S}\cdot {\vec n}_{\perp}$. The inequality $\xi ^2<1$ indicates that the system is spin squeezed.

To find the relation between spin squeezing and quantum entanglement, we first give the following lemma:\\
{\bf Lemma 1}: {\em For a symmetric separable state of $N$ qubits, the correlation function
$\langle\sigma_{i{\vec n}_\perp}\otimes\sigma_{j{\vec n}_\perp}\rangle \ge 0$, where $i$ and $j$ can take any values from 1 to $N$ as long as they are different, and $\sigma_{i{\vec n}_\perp}={\vec \sigma}_i\cdot {\vec n}_\perp$.} \\
{\it Proof}: We first note that the expectation values $\langle\sigma_{i{\vec n}_\perp}\rangle$ and the correlation function $\langle\sigma_{i{\vec n}_\perp}\otimes\sigma_{j{\vec n}_\perp}\rangle\; \forall i\neq j$ are independent of indices due to the exchange symmetry.
The symmetric separable state is given by 
\begin{equation}
\rho_{\text{sep}}=\sum_{k}p_k\rho^{(k)}\otimes\rho^{(k)}\otimes\cdots\otimes\rho^{(k)}
\end{equation} 
with $\sum_{k}p_k=1$. The correlation function $\langle\sigma_{i{\vec n}_\perp}\otimes\sigma_{j{\vec n}_\perp}\rangle$ over the separable state can be obtained from the two-qubit reduced density matrix 
\begin{equation}
\rho_{ij}=\text{Tr}_{\{1,2,\ldots,N\}\backslash \{i,j\}}(\rho_{\text{sep}})=\sum_kp_k\rho^{(k)}\otimes\rho^{(k)},
\end{equation}
yielding
\begin{align}
\langle\sigma_{i{\vec n}_\perp}\otimes\sigma_{j{\vec n}_\perp}\rangle=&\sum_kp_k \text{Tr}_{ij}
[(\rho^{(k)}\otimes\rho^{(k)})(\sigma_{i{\vec n}_\perp}\otimes\sigma_{j{\vec n}_\perp})]\nonumber\\
=&\sum_{k}p_k
\langle\sigma^{(k)}_{i{\vec n}_\perp}\rangle
\langle\sigma^{(k)}_{j{\vec n}_\perp}\rangle\nonumber\\
=&\sum_{k}p_k
\langle\sigma^{(k)}_{i{\vec n}_\perp}\rangle^2\ge 0. \quad \Box
\end{align}

From Lemma 1, we immediately have\\
{\bf Proposition 1}: {\it For an arbitrary symmetric multiqubit state, 
spin squeezing implies pairwise entanglement}.\\
{\it Proof}: Due to the exchange symmetry we may write the expectation value $\langle S_{\vec{n}_\perp}^2\rangle$ as
\begin{equation}
\langle S_{\vec{n}_\perp}^2\rangle=\frac{1}{4}[N+N(N-1)\langle\sigma_{i{\vec n}_\perp}\otimes\sigma_{j{\vec n}_\perp}\rangle]. 
\end{equation}
Substituting the above equation into \rf{eq:squeeze} leads to
\begin{equation}
\xi^2=\frac{4\langle S_{{\vec n}_\perp}^2\rangle}{N}=
1+(N-1)
\langle\sigma_{i{\vec n}_\perp}\otimes\sigma_{j{\vec n}_\perp}\rangle. 
\end{equation}
The above equation shows that spin squeezing is equivalent to negative pairwise correlation ($\langle\sigma_{i{\vec n}_\perp}\otimes\sigma_{j{\vec n}_\perp}\rangle<0$)~\cite{Kitagawa2}. 
This equivalence relation and the above lemma directly leads to the proposition.
$\Box$

Having shown the close relation between spin squeezing and pairwise entanglement, we now proceed to give a quantitative relation between the squeezing parameter and the concurrence~\cite{Conc}. 
We consider an even (odd) pure or mixed state $\rho$. The even (odd) state refers to the state for which only
Dicke states~\cite{Dicke} $|n\rangle_J\equiv |J,-J+n\rangle$ with even (odd) $n$ contribute. For examples, the pure even and odd states are given by
\begin{equation}
|\Psi\rangle_{\text{e}}=\sum_{\text{even n}}c_n|n\rangle_J, \quad |\Psi\rangle_{\text{o}}=\sum_{\text{odd n}}c_n|n\rangle_J,
\end{equation}
respectively. As we will see in the next section, these states can be dynamically generated via a large class of Hamiltonians, and can also be obtained as a superposition of spin coherent states~\cite{WangSuper}.

For the even and odd states, we immediately have the following property
\begin{equation}
\langle S_\beta \rangle =\langle S_zS_\beta \rangle =\langle S_\beta
S_z\rangle =0,\quad \beta \in \{x,y\}.  \label{jxy}
\end{equation}
Therefore, the mean spin is along the $z$ direction. We assume that 
the mean spin satisfies $\langle S_z\rangle\neq 0$.

With the mean spin along the $z$ direction, we have ${\vec n}_\perp=(\cos\theta,\sin\theta,0)$, and thus the operator $S_{{\vec n}_\perp }$ can be written as 
\begin{equation}
S_{\theta}={\vec S}\cdot{\vec n}_\perp=\cos \theta S_x+\sin \theta S_y. 
\end{equation}
So, the squeezing parameter becomes 
\begin{align}
\xi ^2 =&\frac 4N\min_{\theta} \langle S_{\theta}^2\rangle \nonumber\\
=&\frac 2N\min_\theta [\langle S_x^2+S_y^2\rangle 
+\cos (2\theta )\langle S_x^2-S_y^2\rangle  \nonumber\\
&+\sin (2\theta )\langle [S_x,S_y]_+\rangle ]  \nonumber \\
=&\frac 2N[\langle S_x^2+S_y^2\rangle -\sqrt{\langle S_x^2-S_y^2\rangle
^2+\langle [S_x,S_y]_+\rangle ^2}  \nonumber \\
=&1+\frac N2-\frac 2N[\langle S_z^2\rangle +|\langle S_{+}^2\rangle |],
\label{xi}
\end{align}
where $S_\pm=S_x\pm iS_y$ are the ladder operators, and $[A,B]_{+}=AB+BA$ is the anticommutator for operators $A$ and $B.$

From Eq.~\rf{xi}, we see that the squeezing parameter is determined by a sum of two expectation values 
$\langle S_z^2\rangle $ and $\langle S_{+}^2\rangle ,$ and hence the
calculations are greatly simplified. The larger the sum the deeper the spin squeezing. 
We also see that the squeezing
parameter is invariant under rotation along the $z$ direction, i.e., the
squeezing parameter for $\rho $ is the same as that for
$e^{-i\theta S_z}\rho e^{i\theta S_z}$. 

Since the inequality $\langle S_z^2\rangle \leq N^2/4$ always holds, we obtain a lower bound for the squeezing parameter
\begin{equation}
\xi^2\ge 1-\frac{2}{N}
|\langle S_+^2\rangle|.
\end{equation} 
From the above equation, we read that if $|\langle S_{+}^2\rangle|
=0, $ then the squeezing parameter $\xi ^2\geq 1$, which implies a necessary condition for spin squeezing of even and odd states is $|\langle S_+^2\rangle|\neq 0$.
A direct consequence of this necessary condition is that the
Dicke state $|n\rangle_J$ exhibits no spin squeezing since $|\langle S_{+}^2\rangle|=0$. The associated squeezing parameter is given by 
\begin{equation}
\xi^2=1+\frac{2n(N-n)}{N}\ge 1. 
\end{equation}
However, Dicke states can be pairwise entangled~\cite{WangKlaus} even though they are not spin squeezed.

Spin squeezing is related to pairwise correlations
, and negative pairwise correlation is equivalent to 
spin squeezing~\cite{Kitagawa2}. Then, for our even and odd states, we have\\
{\bf Proposition} 2: {\em A necessary and sufficient condition for spin squeezing of even and odd states is given by
\begin{equation}
|u|-y=|\langle \sigma _{i+}\otimes\sigma _{j+}\rangle |+
\frac{\langle \sigma _{iz}\otimes\sigma_{jz}\rangle}{4} -\frac{1}{4}>0.  \label{cond1}
\end{equation}
where
\begin{equation}
u=\langle \sigma _{i+}\otimes\sigma _{j+}\rangle,\quad y =\frac 14\left( 1-\langle \sigma _{iz}\otimes\sigma _{jz}\rangle \right). \label{uy}
\end{equation}
}\\
{\it Proof}: By considering the exchange symmetry,  we have
\begin{equation}
\langle S_{+}^2\rangle =N(N-1)u, \quad \langle S_z^2\rangle =\frac {N^2}4-{N(N-1)}y.
\end{equation}
Substituting the above equation into Eq.~(\ref{xi}) we rewrite
the squeezing parameter as
\begin{align}
\xi ^2=&1-2(N-1)(|u|-y)  \nonumber\\
=&1-2{(N-1)}\left[|\langle \sigma _{i+}\otimes\sigma _{j+}\rangle |+
\frac{\langle\sigma _{iz}\otimes\sigma _{jz}\rangle}{4} -\frac{1}{4}\right]. \label{xi2xi}
\end{align}
We see that spin squeezing is determined by the two
correlation functions $\langle \sigma _{iz}\otimes\sigma _{jz}\rangle $ and $%
\langle \sigma _{i+}\otimes\sigma _{j+}\rangle$.
{}From Eq.~(\ref{xi2xi}), we obtain the proposition. $\Box$

The two correlation functions $\langle \sigma _{iz}\otimes\sigma _{jz}\rangle $ and $\langle \sigma _{i+}\otimes\sigma _{j+}\rangle $ can be obtained from the reduced density matrix $\rho _{ij}=$Tr$_{\{1,2,...,N\}\backslash \{i,j\}}(\rho )$. The reduced density matrix with the exchange symmetry 
is given by~\cite{WangKlaus} 
\begin{equation}
\rho _{ij}=\left( 
\begin{array}{llll}
v_{+} & x_{+}^{*} & x_{+}^{*} & u^{*} \\ 
x_{+} & y & y & x_{-}^{*} \\ 
x_{+} & y & y & x_{-}^{*} \\ 
u & x_{-} & x_{-} & v_{-}
\end{array}
\right),  \label{eq:rhogood}
\end{equation}
in the standard basis $\{|00\rangle ,|01\rangle,|10\rangle
,|11\rangle \}$. The following lemma on the reduced density matrix is useful for later discussions:\\
{\bf Lemma 2}: {\it The matrix elements of $\rho_{ij}$ can be determined by
\begin{align}
v_{\pm } =&\frac{N^2-2N+4\langle S_z^2\rangle \pm 4\langle S_z\rangle (N-1)%
}{4N(N-1)},  \nonumber  \\
x_{\pm } =&\frac{(N-1)\langle S_{+}\rangle \pm \langle
[S_{+},S_z]_{+}\rangle }{2N(N-1)},  \nonumber \\
y =&\frac{N^2-4\langle S_z^2\rangle }{4N(N-1)}, \quad u =\frac{\langle S_{+}^2\rangle }{N(N-1)}.  \label{eq:bbbaaa}
\end{align}
}\\
{\it Proof}: The matrix elements can be represented by expectation values of Pauli spin operators of the two qubits. $v_\pm$ and $x_\pm$ are given  by
\begin{align}
v_{\pm } =&\frac 14\left( 1\pm 2\langle \sigma _{iz}\rangle +\langle \sigma
_{iz}\otimes\sigma _{jz}\rangle \right) ,  \nonumber \\
x_{\pm } =&\frac 12(\langle \sigma _{i+}\rangle \pm \langle \sigma
_{i+}\otimes\sigma _{jz}\rangle ),
\label{eq:para1}
\end{align}
and $u$ and $y$ are given by Eq.~(\ref{uy}).

Due to the exchange symmetry, we have 
\begin{align}
&\langle \sigma _{i\alpha }\rangle =\frac{2\langle S_\alpha \rangle }N%
,\; \langle \sigma _{i+}\rangle =\frac{\langle S_{+}\rangle }N,  \;\langle \sigma _{i\alpha }\sigma _{j\alpha }\rangle =\frac{4\langle
S_\alpha ^2\rangle -N}{N(N-1)}, \nonumber\\
&\langle \sigma _{ix}\sigma _{jy}\rangle =\frac{2\langle
[S_x,S_y]_{+}\rangle }{N(N-1)},  \;
\langle \sigma _{i+}\sigma _{jz}\rangle =\frac{\langle
[S_{+},S_z]_{+}\rangle }{N(N-1)}.  \label{eq:relation}
\end{align}
From Eqs.~(\ref{eq:para1}) and (\ref{eq:relation}), we may thus express the
matrix elements of $\rho _{12}$ in terms of the expectation values
of the collective operators. $\Box$ 

The concurrence quantifying the entanglement of a pair of qubits can be calculated from the reduced density matrix. 
It is defined as~\cite{Conc} 
\begin{equation}
{\cal C}=\lambda _1-\lambda _2-\lambda _3-\lambda _4,
\label{Cdef}
\end{equation}
where the quantities $\lambda _i$ are the square roots of the eigenvalues 
in descending order of
the matrix product 
\begin{equation}
\varrho _{12}=\rho _{12}(\sigma _{1y}\otimes \sigma _{2y})\rho
_{12}^{*}(\sigma _{1y}\otimes \sigma _{2y}).  \label{varrho}
\end{equation}
In (\ref{varrho}), $\rho _{12}^{*}$ denotes the complex
conjugate of $\rho _{12}$. Note that we did not use the max function in the above definition of the concurrence~\cite{Conc}. Therefore, the negative concurrence implies no entanglement here.

Both the squeezing parameter and the concurrence are determined by some correlation functions. So, they may be related to each other. The quantitative relation is given by \\
{\bf Proposition 3}: {\em If} $\xi^2\le 1$ ($|u|\ge y$) {\em for even and odd states, then}
\begin{equation}
\xi ^2=1-(N-1){\cal C}\text{.} \label{main}
\end{equation}\\
{\it Proof}: For our state $\rho$, from Eq.~(\ref{jxy}) and Lemma 2,
it is found that $x_{\pm }=0.$ Therefore, the reduced density matrix becomes 
\begin{equation}
\rho _{ij}=\left( 
\begin{array}{llll}
v_{+} & 0 & 0 & u^{*} \\ 
0 & y & y & 0 \\ 
0 & y & y & 0 \\ 
u & 0 & 0 & v_{-}
\end{array}
\right).  \label{eq:rhogoodg}
\end{equation}

For this reduced density matrix~(\ref{eq:rhogoodg}), the associated concurrence is given by~\cite{WangKlaus}
\begin{equation}
{\cal C}=\left\{ 
\begin{array}{l}
2(|u|-y),\quad\quad\text{ if }2y\le\sqrt{v_{+}v_{-}}+|u|; \\ 
2(y-\sqrt{v_{+}v_{-}}),\text{ if }2y> \sqrt{v_{+}v_{-}}+|u|.
\end{array}
\right.  \label{eq:c2}
\end{equation}
If $|u|\ge y$, we have 
\begin{equation}
2y\leq 2|u|\leq |u|+\sqrt{v_{+}v_{-}},
\end{equation} 
where we have used the fact 
\begin{equation}
v_+v_-\ge |u|^2,\quad v_\pm\ge 0.
\end{equation}
Then, the concurrence \rf{eq:c2} simplifies to 
\begin{equation}
{\cal C}=2(|u|-y).  \label{c3}
\end{equation}
By comparing Eqs.~(\ref{xi2xi}) and (\ref{c3}), we obtain the proposition. $\Box$

According to Proposition 3, we have 
\begin{equation}
{\cal C}=\left\{ 
\begin{array}{ll}
0                & \text{if} ~~\xi^2=1 \\ 
\frac{1-\xi^2}{N-1}>0 & \text{if} ~~\xi^2<1,
\end{array}
\right.   \label{eq:cxxx}
\end{equation}
from which we read that (i) if the squeezing parameter $\xi^2=1$ (no squeezing) for even and odd states, then the concurrence is zero (no entanglement); and (ii) if $\xi^2<1$, there is squeezing, then we have a one-to-one relation between the spin squeezing and pairwise entanglement. However, for the case of $\xi^2>1$,  the concurrence can be positive, and we cannot have ${\cal C}<0$ as exemplified earlier by the Dicke states (Dicke states are simplest cases of even and odd states).  Although the squeezing parameter $\xi^2>1$ implies ${\cal C}<0$ is not valid in general, in the next section we will observe that for some even and odd states the squeezing parameter $\xi^2>1$ does imply ${\cal C}<0$, thereby establishing an equivalence between pairwise entanglement and spin squeezing.

\section{Hamiltonian evolution}

Now we consider a class of states dynamically generated from $|0\rangle_J$ via the following Hamiltonian
\begin{equation}
H =\mu S_x^2+\chi S_y^2+\gamma(S_xS_y+S_yS_x)+f(S_z)
\end{equation}
with $f$ being a function of $S_z$. 
When 
\begin{equation}
\chi=\gamma=f(S_z)=0 
\end{equation}
and 
\begin{equation}
\mu=\chi=f(S_z)=0, 
\end{equation}
the Hamiltonian reduces to the one-axis twisting Hamiltonian~\cite{Barry89,Kitagawa} and the two-axis countertwisting Hamiltonian~\cite{Kitagawa}, respectively. 
When 
\begin{equation}
\chi=\gamma=0, \quad f(S_z)=\Omega S_z, 
\end{equation}
Hamiltonian $H$ reduces to the one considered
in Refs.~\cite{Milburn,Law,AndersKlaus,Leggett,Hines,Pu1}, namely, the one-axis twisting Hamiltonian with a transverse field.
The one-axis twisting Hamiltonian~\cite{Kitagawa} may be realized in various quantum systems including quantum optical systems~\cite{Barry89}, ion traps~ \cite{ions}, quantum dots~\cite{Dot}, cavity quantum electromagnetic dynamics~\cite{Qed}, liquid-state nuclear magnetic resonance (NMR)  system~\cite{Nmr}, and Bose-Einstein condensates~\cite{Sorensen01,uffe}. Experimentally, it has been implemented to produce four-qubit maximally entangled states in an ion trap~\cite{sackett}.

The Hamiltonian exhibits a parity symmetry,
\begin{equation}
[e^{i\pi S_z},H]=[(-1)^{\cal N},H]=0,  \label{symmetry}
\end{equation}
where ${\cal N}=S_z+J$ is the `number operator' of the system, and $(-1)^{\cal N}$ is the parity operator. In other words, the Hamiltonian is invariant under  $\pi$ rotation about the $z$ axis. The symmetry can be easily seen from the transformation 
\begin{equation}
e^{i\pi S_z}(S_x,S_y,S_z)e^{-i\pi S_z}=(-S_x,-S_y,S_z).
\end{equation}

We assume that the initial density operator is chosen to be
\begin{equation}
\rho(0)=|0\rangle_J\langle 0|, 
\end{equation}
where $|0\rangle_J=|1\rangle \otimes
|1\rangle \otimes \cdots \otimes |1\rangle$, and state $|1\rangle $
denotes the ground state of a qubit. 
The density operator at time $t$ is
then formally written as 
\begin{equation}
\rho(t) =e^{-iHt}\rho(0)e^{iHt}.
\end{equation}

The parity symmetry of $H$ in (\ref{symmetry}) leads to the useful property 
given by Eq.~(\ref{jxy}). For example,
\begin{align}
\langle S_x\rangle 
=&\text{Tr}[S_xe^{-iHt}\rho(0)e^{iHt}]
\nonumber\\
=&\text{Tr}[S_xe^{-iHt}e^{-i\pi S_z}\rho(0)e^{i\pi S_z}e^{iHt}]\nonumber\\
=&\text{Tr}[e^{i\pi S_z}S_xe^{-i\pi S_z}\rho(t)]\nonumber\\
=&-\langle S_x\rangle.
\end{align}
From another point view, the state $\rho(t)$ is an even state since the Hamiltonian is quadratic in generators $S_x$ and $S_y$ and the initial state is an even state. Then, Eq.~(\ref{jxy}) follows directly. 
Since state $\rho(t)$ is an even state, we may apply the results in the last section. Next, we consider three representative model Hamiltonians for generating spin squeezing, which are special cases of Hamiltonian $H$.

\subsection{One-axis twisting Hamiltonian}
We first examine the well-established one-axis twisting model
~\cite{Barry89,Kitagawa},  
\begin{equation}
H_1=\mu S_x^2,
\end{equation}
for which we have \\
{\bf Lemma 3}: {\em For the state dynamically  generated from $|0\rangle_J$ via the one-axis twisting Hamiltonian, we always have $\xi^2\le 1$.}\\
{\it Proof}: From the results of Refs~\cite{Kitagawa,WangKlaus}, we have
the following expectation values~($\bar{\mu} =2\mu t$) 
\begin{align}
\langle S_x^2\rangle  =&N/4, \nonumber  \\
\langle S_y^2\rangle  =&\frac 18\left( N^2+N-N(N-1)\cos ^{N-2}\bar{\mu} \right), 
\nonumber \\
\langle S_z^2\rangle  =&\frac 18\left( N^2+N+N(N-1)\cos ^{N-2}\bar{\mu} \right). 
\end{align}
Then, we obtain a useful relation for density operator $\rho(t)$ at any time $t$,
\begin{equation}
\langle S_x^2-S_y^2\rangle =\langle S_z^2\rangle -N^2/4=-N(N-1)y, 
\label{eq:bbbb}
\end{equation}
where we have used Eq.~(\ref{eq:bbbaaa}). From the above equation, we obtain
\begin{align}
|u|^2 =&
\frac 1{N^2(N-1)^2}(\langle S_x^2-S_y^2\rangle ^2+\langle
[S_x,S_y]_{+}\rangle ^2) \nonumber\\
\geq &\frac{\langle S_x^2-S_y^2\rangle ^2}{N^2(N-1)^2}\nonumber\\
=&y^2,
\end{align}
which implies $|u|\geq y$ at any time (note that $y\ge 0$). Therefore, the squeezing parameter always satisfies $\xi^2\le 1$. $\Box$

Then, from Proposition 3 and Lemma 3, we obtain \\ 
{\bf Proposition 4}: {\em For the state dynamically  generated from $|0\rangle_J$ via the one-axis twisting Hamiltonian, it is spin squeezed if and only if it is pairwise entangled. Hence spin squeezing and pairwise entanglement
are equivalent for such state.}

At times for which ${\cal C}=0,$ the state vector is either a product
state or an $N$-partite ($N\ge 3$) maximally entangled state~\cite{ions,sackett} which has no pairwise entanglement, and thus no spin squeezing. 

\subsection{One-axis twisting Hamiltonian with a transverse field}
We consider the one-axis twisting model with an external transverse field described by the Hamiltonian 
\begin{equation}
H_2=\mu S_x^2+\Omega S_z, 
\end{equation}
where $\Omega>0$ is the strength of the transverse field. In general, this model cannot be solved analytically. Numerical results show that the squeezing parameter $\xi^2\le 1$ for the dynamically generated state $\exp(-iH_2t)|0\rangle_J$~\cite{Law}. 
We perform numerical calculations for $N$ from 2 to $100$ qubits, different values of $\Omega$, and $\mu=1$, which indeed display the inequality $\xi^2\le 1$. Therefore, according to Proposition 3, these numerical results suggest that  spin squeezing implies pairwise entanglement and vice versa for such states generated from $|0\rangle_J$ via Hamitonian $H_2$. In the limit of $\Omega\rightarrow 0$, the result of this subsection, of course, reduces to that of the previous one. 

\begin{figure}
\includegraphics[width=0.45\textwidth]{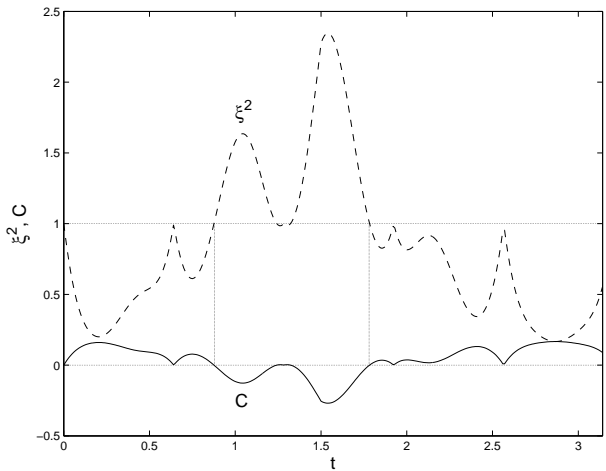}
\caption{\label{entfig}
The spin squeezing parameter and the concurrence against time $t$ for six qubits. The parameter $\gamma$ is chose to be 1.}
\end{figure}

\subsection{Two-axis counter-twisting Hamiltonian}
Finally, we examine the two-axis counter-twisting model described by Hamiltonian 
\begin{equation}
H_3=\frac{\gamma}{2i}(S_+^2-S_-^2). 
\end{equation}
For the state generated from $|0\rangle_J$ via Hamiltonian $H_3$, the squeezing parameter can be larger than 1. Numerical results for $N$ from 2 to 100 and $\gamma=1$ suggest that the relation (\ref{main}) holds for even $N$,
\begin{equation}
{\cal C}=\left\{ 
\begin{array}{ll}
0 & \text{if} ~~\xi^2=1, \\ 
\frac{1-\xi^2}{N-1}>0 & \text{if} ~~\xi^2<1, \\
\frac{1-\xi^2}{N-1}<0 & \text{if} ~~\xi^2>1. 
\end{array}
\right.   \label{eq:cxxxxxx}
\end{equation}
The above equation displays an equivalence relation between spin squeezing and pairwise entanglement for states generated from $|0\rangle_J$ via Hamiltonian $H_3$ with even $N$. The case of $N=6$ is demonstrated
in Fig.~1, which are plots of the spin squeezing parameter and the concurrence against time $t$. We make a conjecture that the spin squeezing and pairwise entanglement are equivalent for the states generated via the one-axis twisting Hamiltonian with an external transverse field for any number $N\ge 2$ or via the two-axis counter-twisting Hamiltonian for any even number of qubits.   

\section{Conclusions}
In conclusion, we have shown that spin squeezing implies pairwise entanglement for arbitrary symmetric multiqubit states. We have identified a large class of multiqubit states, i.e., the even and odd states, for which 
the quantitative relation of the spin squeezing parameter and the concurrence is given. We have proved that spin squeezing implies pairwise entanglement and vice versa for the states generated from $|0\rangle_J$ via the one-axis twisting Hamiltonian. For the states dynamically generated from $|0\rangle_J$ via the one-axis twisting Hamiltonian with a transverse field for any $N\ge 2$ and the two-axis counter-twisting Hamiltonian with any even $N$, numerical results suggest that spin squeezing implies pairwise entanglement and vice versa. As these three model Hamiltonians have been realized in many physical systems, the close relations between the spin squeezing and pairwise entanglement are meaningful and help to understand quantum correlations in these systems.

\acknowledgments
We acknowledge the helpful discussions 
with A. R. Usha Devi, Dominic W. Berry, G\"{u}nter Mahler, and Leigh T. Stephenson. This project has been supported by an Australian Research Council Large Grant.

\end{document}